\documentclass[aip,showpacs,preprintnumbers,amsmath,amssymb]{revtex4-1}

\usepackage{graphicx}
\usepackage{dcolumn}
\usepackage{bm}

\begin{document}

\title{Optical properties of cubic and rhombohedral GeTe}

\author{David J. Singh}

\affiliation{Materials Science and Technology Division,
Oak Ridge National Laboratory, Oak Ridge, Tennessee 37831-6056}

\date{\today}

\begin{abstract}
Calculations of the optical properties of GeTe in the cubic NaCl
and rhombohedral ferroelectric structures are reported. The rhombohedral
ferroelectric distortion increases the band gap from 0.11 eV to 0.38 eV.
Remarkably, substantial changes in optical properties are found even at high
energies up to 5 eV. The results are discussed in relation to the
bonding of GeTe and to phase change materials based on it.
\end{abstract}


\maketitle

\section{Introduction}

GeTe is a well known ferroelectric material with a Curie temperature
$T_C$ of $\sim$670 K depending strongly on sample quality.
\cite{schubert1,schubert2,bierly,goldak,littlewood,pawley}
The related compound, SnTe, is a low temperature ferroelectric
again with strong sensitivity to sample quality. \cite{pawley,iizumi}
PbTe is cubic but near a ferroelectric instability, which
may be important in relation to its thermoelectric properties.
\cite{an,delaire}
Above $T_C$, GeTe has a cubic NaCl type crystal structure, which
undergoes a distortion consisting of a relative displacement of the
Ge and Te sub-lattices along [111] accompanied by a sizable rhombohedral
strain when cooled through $T_C$. At low temperatures this strain
amounts to $\sim$2$^\circ$.
The driving mechanism for this instability was discussed by
Lucovsky and White \cite{luc2} and by
Littlewood \cite{littlewood} in terms of bonding, specifically the
presence of
p-electron bonding between Ge and Te,
though from different perspectives.
Lucovsky and White coined the term ``resonance bonding"
for this type of material, where the high symmetry crystal structure
is not well adapted to the covalent bonding pattern of the elements and
the result is a superposition of different covalent bond networks.
\cite{luc2}

In any case, GeTe has been the subject of renewed interest due to
its relationship with phase change materials used in data storage.
\cite{yamada,yamada2,wuttig,raoux}
The ferroelectricity, as mentioned, is sensitive to disorder.
When alloyed with Sb$_2$Te$_3$, the material adopts a cubic
NaCl type crystalline structure in which one sub-lattice
contains Te and the other contains an approximately charge balanced mixture
of Ge, Sb and vacancies, such that there is ideally one compensating
vacancy ($V$) for each two Sb atoms in the alloy,
i.e. an approximate formula (Ge$_{1-3x}$Sb$_{2x}V_x$)Te.
Recently developed optical media (e.g. ``blu-ray") use relatively
low concentrations of Sb, in some cases less than 10\%
(a typical composition is Ge$_8$Sb$_2$Te$_{11}$).
\cite{yamada-spie}

The NaCl structure crystalline state is either
metallic or has a low band gap, $\sim$0.1 eV or less depending on the
detailed composition. Rapid quenching from the melt, however,
yields a dense amorphous phase with a wider gap $\sim$0.5 eV and rather
different properties, specifically a lower bulk reflectivity and much
higher resistivity. This amorphous phase is highly
unusual in that it can
be extremely rapidly converted to the crystalline phase by annealing,
with reported time scales of 10 -- 100 ns (20 ns is typical in practical
application). Furthermore, even shorter crystallization times may be
possible with selected starting conditions. \cite{loke}
While practical materials involve alloying with Sb or
other modifications, similar
behavior is also found in nominally
pure GeTe films, including an amorphous phase
with lower bulk reflectivity than the crystalline phase and rapid,
sub-100 ns crystallization.

Here we report calculations of the optical properties of cubic
and rhombohedral ferroelectric GeTe.
As may be expected there is a change in the band gap between these
states, amounting to $\sim$ 0.3 eV. In addition, we find that
there is a substantial change in optical properties, specifically a
strong decrease in
reflectivity over a wide energy range up to almost 5 eV, or almost
15 times the band gap.
This is a consequence of the role of covalent bonding in the
ferroelectric instability of GeTe.
We discuss the results in the context of phase change materials.
Specifically, based on the results and consideration of existing
experimental data we argue that there may be a relationship between
the amorphous phase and the ferroelectric phase, and in particular
that the amorphous phase may be a form of relaxor ferroelectric,
with implications for the properties of the amorphous phase.

\section{Computational Approach}

The calculations were performed within
density functional theory with a modified
exchange correlation potential designed to reproduce band gaps.
They were done using the general potential linearized
augmented planewave (LAPW) method with local orbitals,
\cite{singh-lo,singh-book}
as implemented in the WIEN2k code. \cite{wien2k}
LAPW sphere radii of 2.5 bohr and 2.6 bohr were used for Ge and Te,
respectively, and spin-orbit was included, as is needed for this compound.
\cite{herman,tung}
We employed experimental crystal structures, i.e. a cubic lattice parameter of
$a$=5.976 \AA, and rhombohedral ferroelectric structure,
$a$=4.16 \AA{}, $c$=10.68 \AA, $u$=0.4752 (Ge at (0,0,0); Te
at (0,0,$u$), in hexagonal coordinates, where $u$=0.5 is the ideal
value for the NaCl structure and $u$=0.25 would correspond to a
zinc blende tetrahedral structure).

The optical properties were obtained using the optical package of
the WIEN2k code, which uses direct calculation of dipole matrix elements.
Importantly, the modified Becke-Johnson potential developed by
Tran and Blaha, denoted TB-mBJ was employed. \cite{mbj}
This potential functional generally yields much improved band gaps
for simple semiconductors and insulators relative to standard density functionals,
\cite{mbj,singh2,koller,singh1}
which are designed to reproduce total energies and do not generally yield band
gaps in reasonable accord with experiment.
\cite{pw91}
This approach was used previously to calculate optical properties of PbTe
and PbSe, which were found to be in good accord with experimental data,
including the 1 eV -- 5 eV energy range,
\cite{ekuma} which is of importance here.
Calculations for Bi$_4$Ti$_3$O$_{12}$ also showed quantitative agreement
with experiment. \cite{bit}

\section{Electronic Structure and Optical Properties}

\begin{figure}
\includegraphics[width=0.7\columnwidth]{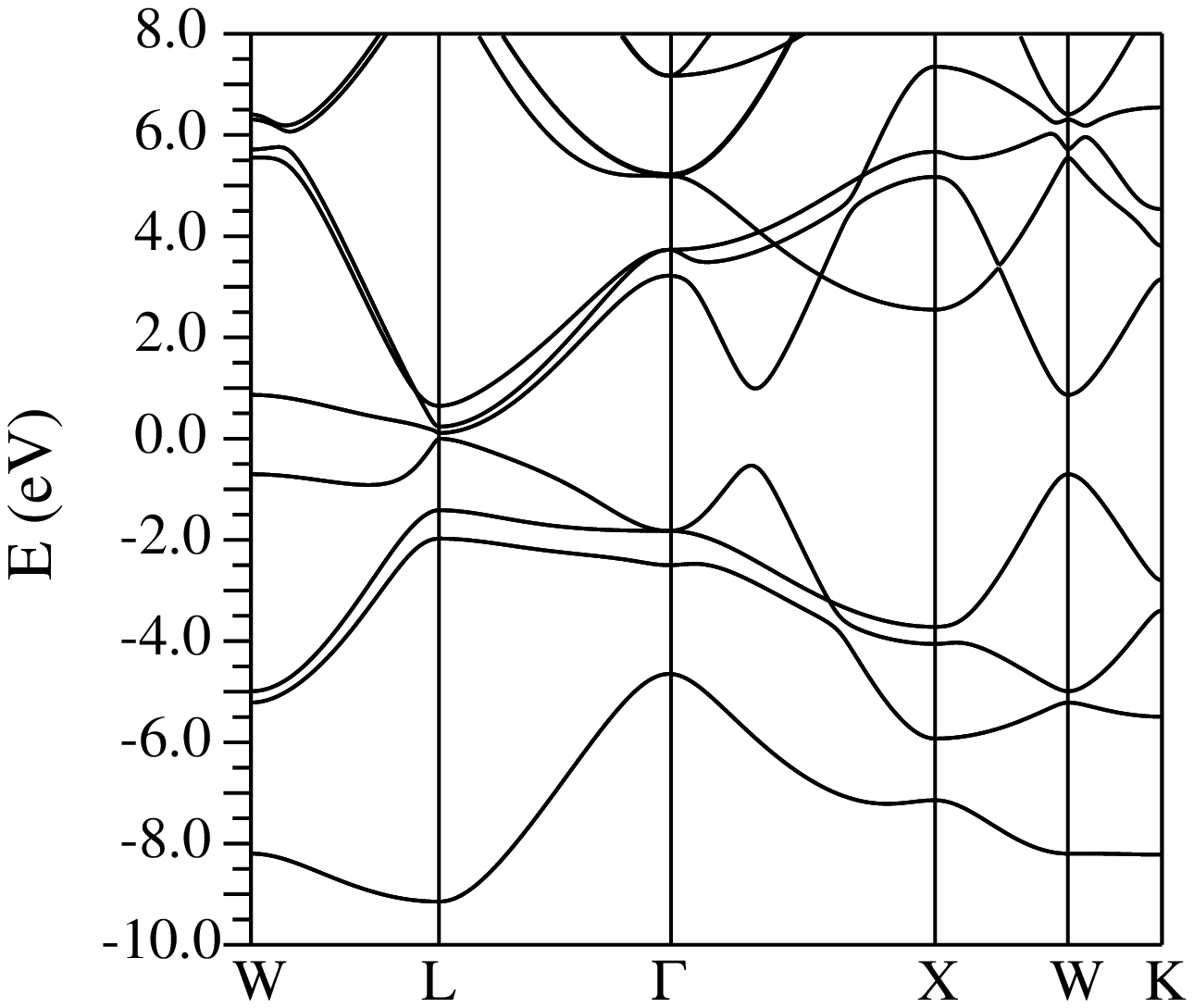}
\includegraphics[width=0.7\columnwidth]{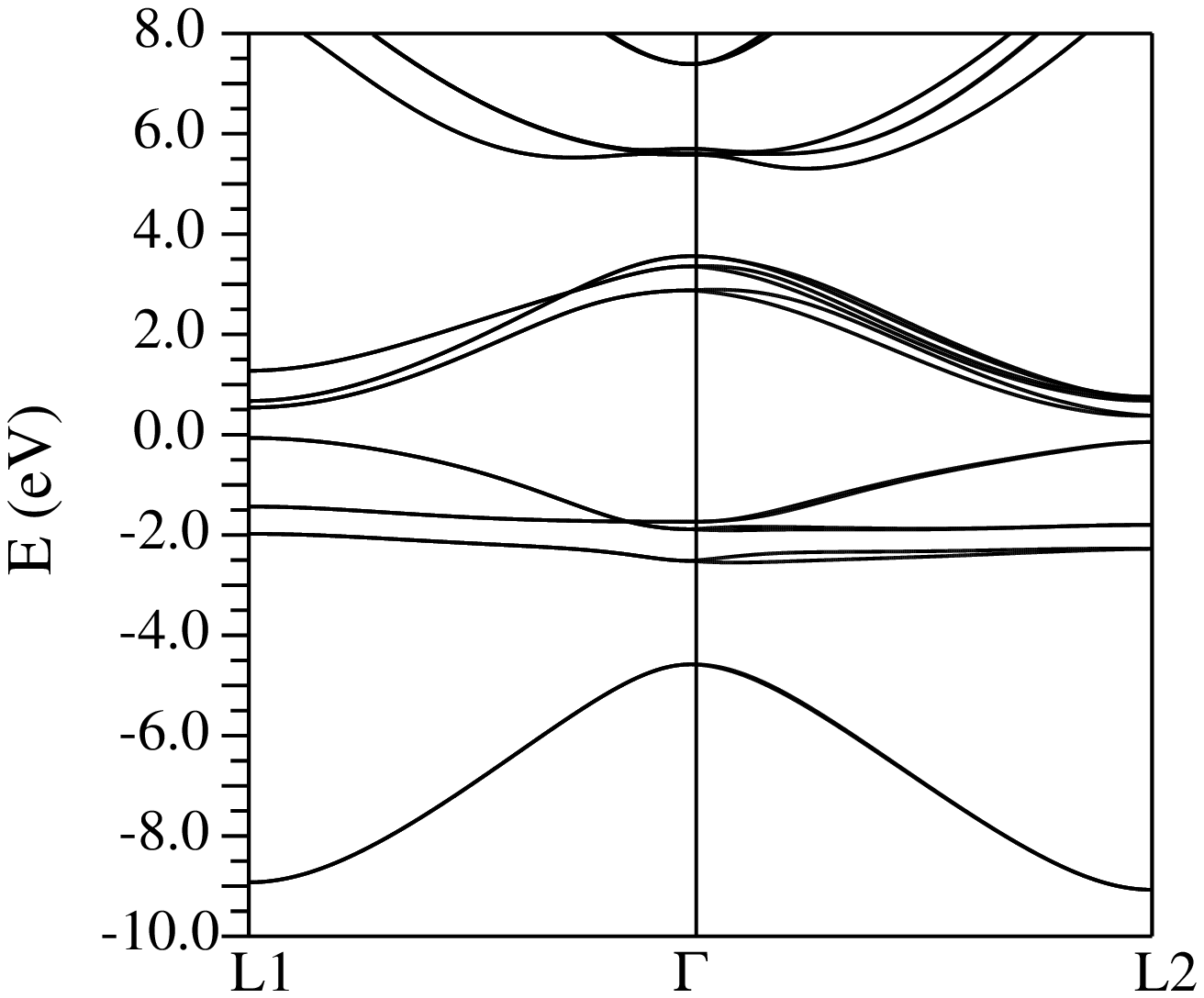}
\caption{Calculated band structure of cubic GeTe (top) and
rhombohedral ferroelectric GeTe showing the non-equivalent
$\Gamma$-L directions  (bottom). The points ``L1" and ``L2"
are the fcc L points along the rhombohedral
$c$-axis and non-$c$-axis directions,respectively.
Note the differences between these two directions, e.g.
in the dispersion of the lowest conduction band from $\Gamma$.
The direct band gaps at ``L1" and ``L2" are 0.61 eV and
0.53 eV respectively, while the indirect gap from the valence
band at ``L1" to the conduction band at ``L2" is 0.45 eV. The lowest
indirect gap is from a valence band away
from a symmetry point
approximately along a line joining an ``L1" to an ``L2" point
to the ``L2" point conduction band and is 0.38 eV.}
\label{bands}
\end{figure}

The relativistic band structure of cubic GeTe has been reported by many authors,
starting with Herman and co-workers and Tung and Cohen. \cite{herman,tung}
Our result is similar to prior reports except for the exact value of the
band gap reflecting the use of the TB-mBJ potential in the present
calculations. Our calculated
gap is 0.11 eV, as shown in Fig.  \ref{bands},
and is direct at the L point of the fcc zone in accord with prior work.
This is also in accord with experimental measurements on NaCl
structure (heavily $p$-type non-ferroelectric) films, which show
a band gap in the range 0.1 eV -- 0.2 eV.
\cite{tsu2,tsu,bahl}
As an aside, we note that this band structure is qualitatively similar
to that of PbTe and PbSe, \cite{singh-pbte,parker}
and is favorable for thermoelectric performance.

Turning to the rhombohedral ferroelectric structure,
according to our calculations, the material has an indirect gap of 0.38 eV
in this state, while the direct gap (which remains at an
L point) is 0.53 eV. The indirect gap in the ferroelectric
state was previously noted by
Yamanaka and co-workers \cite{yamanaka}, and by
Park and co-workers. \cite{park}
While this band gap change is large on a relative scale because of
the small band gap of the cubic phase, it is far below the energy
range where reflectivity changes are seen in phase change materials.
For example, a typical laser probe for ``blu-ray" media is at 405 nm
(3.06 eV). The phase change materials
used in such media show are sizable differences in optical properties
between the NaCl and amorphous structures in this energy range.

\begin{figure}
\includegraphics[height=\columnwidth,angle=270]{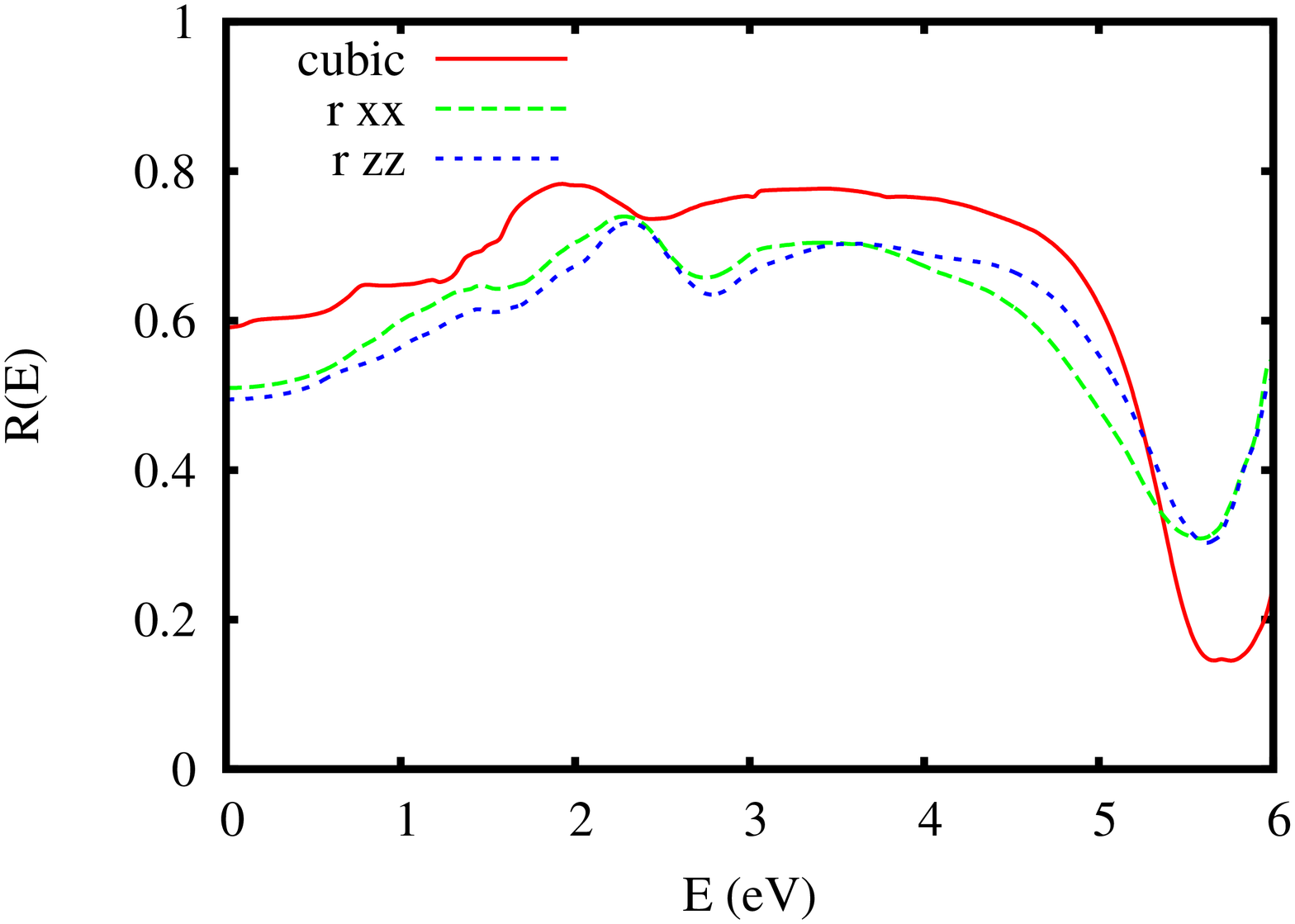}
\caption{Calculated optical reflection for cubic and rhombohedral
ferroelectric GeTe. For the rhombohedral case, ``zz" denotes polarization
along the ferroelectric
polar axis, while ``xx" denotes in polarization perpendicular to it.  }
\label{reflection}
\end{figure}

The calculated bulk optical reflectivity is shown in Fig. \ref{reflection}
(note that the reflectivity of a thin film includes also contributions,
which can be large, from the bottom surface
and multiple reflections depending on the details
of the film thickness and substrate; also the bulk reflectivity is
an upper bound for a bulk sample,
as for a real sample there will also be scattering,
e.g.  due to surface roughness or inhomogeneity,
which will be wavelength dependent
and will reduce the measured reflectivity).
At 3 eV, the calculated direction averaged reflectivity is 
0.68 for the ferroelectric phase and 0.77 for the cubic phase, amounting
to a surprisingly large difference of 13\%.

Park and co-workers reported a detailed study of the optical
properties of GeTe, Sb$_2$Te$_3$ and alloys using ellipsometry measurements
on films as well as density functional calculations. \cite{park}
Yamanaka and co-workers
\cite{yamanaka} presented density functional calculations
of optical constants of GeTe and Ge$_2$Sb$_2$Te$_5$ using the local
density approximation along with experimental
measurements of optical constants.
Optical reflectivity spectra of GeTe over a wide energy range were
reported by Cardona and Greenway on polycrystalline GeTe
films. \cite{cardona}
While the measured
absolute reflectivity is consistently lower than the calculated
values, perhaps because of scattering
in the non-single crystal film,
similar structures are seen, in particular a maximum at $\sim$ 2eV, and a
minimum just above 5 eV. In the data of Cardona and Greenway there is
a small maximum above this latter minimum, while we obtain a larger maximum.
In both the experimental data and the
calculated spectrum there is a smooth fall-off in reflectivity
at energies above this.
The low energy (where off-specular
scattering should be less important)
extrapolated experimental reflectivity is 0.5
as compared to the calculated value 0.59 for cubic GeTe and a direction
averaged 0.5 for ferroelectric, which would constitute good agreement
assuming that the film measured was ferroelectric.

As mentioned, the driving force for the ferroelectric transition
in GeTe has been discussed in terms of $p$-electron bonding.
In the resonant bond picture of Lucovsky and White, \cite{luc2}
electrons in bonding orbitals are shared between
different bonds and both the cubic and ferroelectric states are
substantially covalent (it should be noted that sharing electrons in a bonding
network would normally stabilize a highly covalent high symmetry state as in
aromatic rings and multicenter bonds).
The conclusion that p-electron
bonding is important for the ferroelectricity
has also been reached using first principles 
calculations, though the details of the bonding are
different. \cite{waghmare}
Interestingly, there is a substantial strain
coupling between the frozen-in zone center transverse optic mode
and the rhombohedral strain, \cite{rabe} presumably as a consequence
of the covalent bonding in the ferroelectric state.

\begin{figure}
\includegraphics[height=\columnwidth,angle=270]{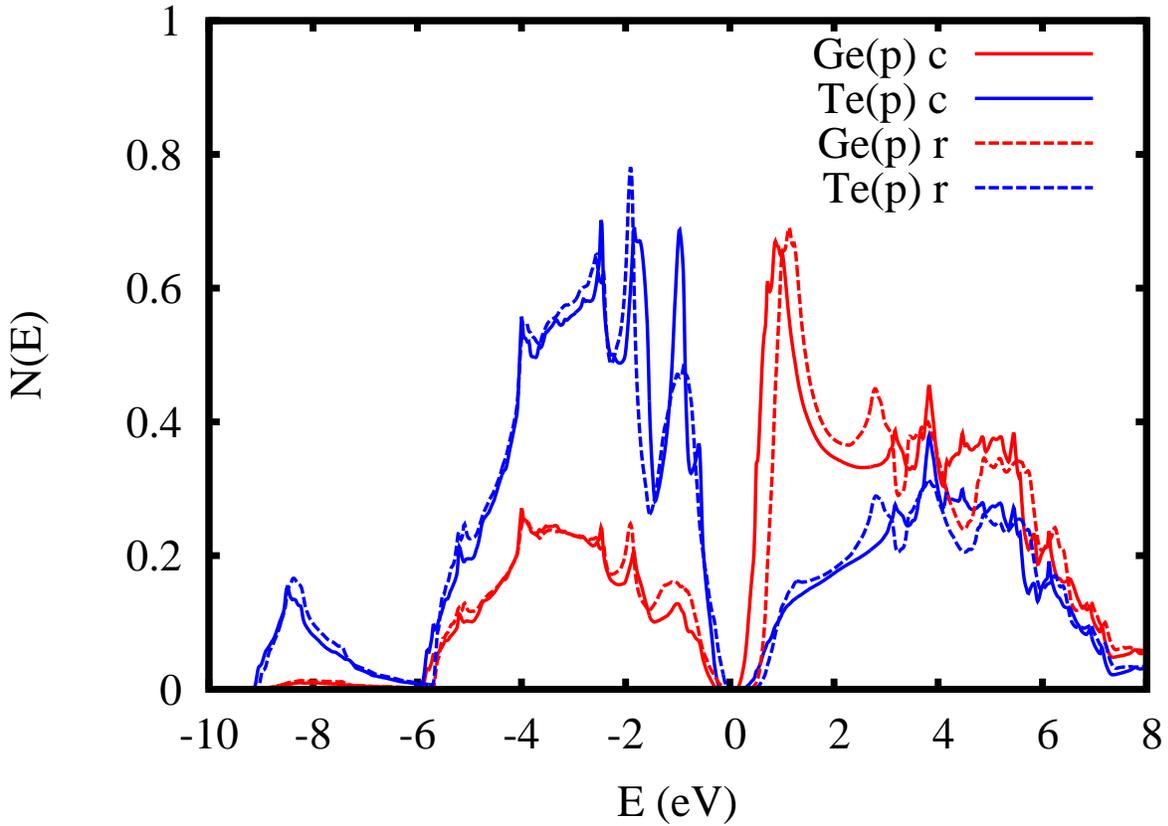}
\caption{Calculated $p$-character
projections of the electronic density of states of NaCl cubic (c)
and rhombohedral (r) GeTe onto the Ge and Te LAPW spheres. The energy
zero is set to the valence band maximum.}
\label{dos}
\end{figure}

Fig. \ref{dos} shows the calculated projections of the electronic density
of states of Ge and Te $p$-character in the LAPW spheres.
This is a quantity that is proportional to the $p$-contributions to the
electronic structure. As may be seen, both the valence and conduction
band regions consist of mixtures of Ge $p$ and Te $p$ states,
similar to what has been shown in prior studies. \cite{waghmare,yamanaka}
Descriptions of bonding in solids are necessarily qualitative
because of the ambiguity of assigning charge and orbital characters
to different sites. Nonetheless, it is apparent from the figure that
the nature of the states in the conduction band is quite different
from those in the valence band, even at the band edges, in contrast to
a simple interpretation of resonant bonds.

Specifically, the valence bands have relatively much higher Te $p$
character than the conduction bands, and conversely for the Ge $p$
character. This is in contradiction to what is anticipated for
a resonant bonded configuration, and is more like what is expected
for an ionic material, consisting nominally of
Ge$^{2+}$ and Te $^{2-}$ ions, with some covalent character arising from
cross gap hybridization between $p$-orbitals on the anion and cation
sites. This is consistent both with the octahedral
coordinated NaCl type crystal structure,
which is an ionic crystal structure, as well as with the results of 
Waghmare and co-workers, \cite{waghmare} who
found highly enhanced Born effective charges,
$Z^*_{\rm Ge}$=-$Z^*_{\rm Te}$=10.8, as compared to the nominal valence of 2.0.
Such large enhancements of the Born effective charge are characteristic
of conventional ferroelectrics such as BaTiO$_3$,
and arise from cross-gap hybridization between nominally occupied states
on the anion (O) and unoccupied states on the active cation (Ti).
\cite{axe,cohen-covalency,zhong-loto}
In contrast, the Born effective charges of covalent semiconductors
are generally reduced from the nominal valences.

Cross-gap hybridization
is understood as a main driving force for ferroelectricity
in conventional oxides such as BaTiO$_3$, \cite{cohen-covalency,zhong-loto}
where the enhanced Born charges reflect an improved bonding with
distortion.
Similarly, the ferroelectricity of GeTe may be regarded as arising from the
instability of an ionic cubic structure due to covalency between
anions and cations.
In view of this mechanism,
one expects changes in the density of states particularly corresponding to
the orbitals involved with the ferroelectric distortion.
In fact, this can be seen in Fig. \ref{dos}, though rather weakly because of the small
displacement of the atoms.
Besides the upward energy shift of the conduction band edge in the
ferroelectric state, one notes an increased $p$ contribution for
the ferroelectric state in the energy range near 3 eV and a decrease
in the vicinity of 4 eV -- 5 eV. This is accompanied by a small downward
shift of the density of states in the valence bands in the range near -2 eV.
While these changes are small they underlie the change in optical
reflectivity at energies well above the band gap.
In short, the fact that the ferroelectric instability is connected
to the $p$-bonding results in large changes in reflectivity between the cubic
and ferroelectric structures.

\section{Discussion and Conclusions}

This may have implications for understanding the phase
change materials.
Experimental investigation and first principles structural studies and
molecular dynamics have shown that the local structure of the
amorphous phase of GeTe-Sb$_2$Te$_3$ alloys
is more characteristic of a $p$-bonded covalent
material than the cubic structure.
\cite{caravati,akola,hegedus,sun,dasilva,
xu,akola1,liu,raty,krbal,kalikka,lee1,skelton,zhang}
One complication is that the amorphous phase is not an equilibrium
thermodynamic phase but is a frozen in structure, which in principle
depends on the sample history.
In any case,
most studies have focused on the composition Ge$_2$Sb$_2$Te$_5$,
which has 20\% vacancies on the cation site. \cite{nonaka}
Not surprisingly,
there are distortions around the vacancies leading
to low coordinated Te atoms, while other
Te atoms retain octahedral coordination. \cite{caravati,sun,akola}
Recent work shows an important role for the vacancy ordering in determining
the electrical resistivity. \cite{zhang}
While all the
coordinations are changed in the amorphous structures, detailed
studies indicate the presence of both of large
numbers of approximately octahedral coordinated atoms, \cite{hegedus}
as in the NaCl structure and the ferroelectric
structure of GeTe, as well
as some, possibly
small, \cite{dasilva,raty}
fraction of approximately tetrahedral sites for Ge.
\cite{akola,liu}

What the present results show is that just the relatively subtle
structural changes associated with ferroelectricity can lead to
large changes in optical properties.
Based on this, one can speculate that this may be an important
aspect of the structure of the amorphous phase as it relates to
optical reflectivity. We observe that the crystallization temperatures
of $\sim$450 K -- 600 K are comparable to the ferroelectric
Curie temperature of GeTe. In general, long range
ferroelectric order can be
suppressed by either disorder or doping.
Disorder in particular can preserve the
local chemical environment that favors off-centering. It may be that
the amorphous state, characterized by disordered, locally inhomogeneous
vacancies, has characteristics of a relaxor ferroelectric,
while the less
disordered crystalline phase is more similar to the non-ferroelectric cubic
state.

Relaxor ferroelectrics, which can be produced from standard ferroelectrics
by strong disorder, are characterized by a Burns temperature, $T_B$, below
which the local structure becomes similar to a ferroelectric,
while the long range symmetry does not. \cite{bokov,egami}
Relaxor ferroelectrics do not show long range ferroelectric order, but
do show characteristics of the ferroelectric phase. In the present case,
because of the existence of nanoscale polar regions (so called polar
nanoregions, or PNR) in relaxor materials below $T_B$, one may
expect optical properties similar to the ferroelectric phase.
Besides their ferroelectric like local structures, characteristics of
relaxor ferroelectrics are an enhanced dielectric response in the
form of a prominent broad peak in the temperature dependent static
dielectric constant, $\epsilon_0$, along with a characteristic
frequency dependence of the peak.
We suggest that it will be of interest to study the dielectric
behavior and other properties of the amorphous phase to determine
whether signatures of relaxor ferroelectricity are present. This
may provide new insight into the structure of the amorphous phase
and additionally
may potentially have implications for the electrical properties
of devices based on electrical sensing such as
phase change random access memories.

\acknowledgments

I am grateful for helpful discussions with R.O. Jones and S.R. Elliott.
Work at ORNL was supported by the Department of Energy, Basic Energy Sciences,
Materials Sciences and Engineering Division.

\bibliography{gete}

\end{document}